# Vortex pinning vs superconducting wire network: origin of periodic oscillations induced by applied magnetic fields in superconducting films with arrays of nanomagnets

A Gomez[1], J del Valle[1], E M Gonzalez[1,2], C E Chiliotte[3], S J Carreira[3], V Bekeris[3], J L Prieto[4], Ivan K Schuller[5] and J L Vicent[1,2]

[1]Departamento de Física de Materiales, Facultad CC. Físicas, Universidad Complutense, 28040 Madrid, Spain.
[2]IMDEA-Nanociencia, Cantoblanco, 28049 Madrid, Spain.
[3]Departamento de Física, Facultad de C. Exactas y Naturales, Universidad de Buenos Aires, Argentina.
[4]ISOM-ETSIT, Universidad Politécnica de Madrid, 28040 Madrid, Spain.
[5]Department of Physics and Center for Advanced Nanoscience (CAN), University of California-San Diego, La Jolla CA 92093, USA.

E-mail: cygnus@ucm.es

**Abstract.** Hybrid magnetic arrays embedded in superconducting films are ideal systems to study the competition between different physical (such as the coherence length) and structural length scales such as available in artificially produced structures. This interplay leads to oscillation in many magnetically dependent superconducting properties such as the critical currents, resistivity and magnetization. These effects are generally analyzed using two distinct models based on vortex pinning or wire network. In this work, we show that for magnetic dot arrays, as opposed to antidot (i.e holes) arrays, vortex pinning is the main mechanism for field induced oscillations in resistance R(H), critical current $I_c(H)$, magnetization M(H) and ac-susceptibility $\chi_{ac}(H)$ in a broad temperature range. Due to the coherence length divergence at $T_c$, a crossover to wire network behaviour is experimentally found. While pinning occurs in a wide temperature range up to $T_c$, wire network behaviour is only present in a very narrow temperature window close to $T_c$. In this temperature interval, contributions from both mechanisms are operational but can be experimentally distinguished.

*Pacs numbers:* 74.25.F-, 74.25.Wx, 74.78. Na, 74.25.Ha

## 1. Introduction

Geometric constrictions are powerful tools used to explore unusual features present in superconductors. In general, two different approaches are adopted: i) weak electric links which connect superconducting areas and ii) wire networks of narrow superconducting tracks. Superconducting constrictions play important roles in many different superconducting areas; such as superconducting electronic devices [1] and high temperature superconductors [2].

Applied magnetic fields may produce interesting effects in nanostructured superconductors, for instance, they may induce different types of periodic responses. As examples, Shapiro steps [3], provide remarkable fingerprints of weak links [4] whereas magnetically induced Little-Parks [5] critical temperature oscillations are a distinct hallmark of wire networks.

A different way to produce constrictions into superconductors is using arrays of non-superconducting materials in superconducting films. The interplay between magnetic arrays and superconducting films could lead to reversible on/off switching of magnetically induced weak links [6]. The main goal for the study of these hybrid systems is to enhance pinning mechanisms or to modify the vortex lattice dynamics [7].

In hybrid superconducting/non-superconducting systems commensurability effects between the vortex lattice and the artificial non-superconducting array generate well defined periodic features in the magnetic [8] and transport [7] properties, at temperatures close to the superconducting critical temperature ($T_c$). This matching effect has been ascribed to enhancement of the vortex lattice pinning by the artificial lattice at matching conditions, i. e. when there are integer number of vortices per plaquette [7]. Similar magnetoresistance minima and ordered magnetic flux structures [9, 10] can be induced, assuming a very different scenario: Little-Parks critical temperature oscillations in a wire network. Recently, Zhang et al. [11] have reported that interstitial vortex state and wire network-like state could be reached increasing the applied magnetic field, at constant temperature, in Nb films with array of large antidots (holes). Earlier, the influence of the size and magnetic state of the dots in the superconductor–normal-metal phase boundary has been reported [12-14]. Moshchalkov *et al.* [15] found that, for large enough antidot size, the wire network regime is always reached by choosing an appropriate combination of temperature and antidot radii. Using Ni dot arrays embedded in Nb films, Hoffmann *et al.* [16] claim features of wire network behavior when the separation of the dots is of the order of the superconducting coherence length. Patel *et al.*[17] have ruled out vortex pinning as the origin of these striking magnetoresistance minima close to critical temperature in superconducting films with arrays of holes. These authors conclude that the magnetoresistance dips are originated from antidot-induced $T_c$ suppression; i. e. the Little-Parks effect governs the magnetoresistance minima. Recently, using the Little-Parks effect only, Latimer *et al.* [18] have studied anisotropic properties of Nb films with different hole arrays. On the other hand, moving vortices are needed to explain magnetoresistance oscillations in superconducting films with antidot arrays [19,20].

In the present work, we have studied Nb films with arrays of magnetic Ni dots with similar dimensions that have been previously reported in the literature [7]. In these hybrid superconductors two types of pinning potentials are present: i) random intrinsic potentials due to the superconducting film defects and ii) periodic potentials due to the magnetic array. The interplay between these two kinds of potentials governs the periodic responses [21-23]. The main features of the commensurability effect, i. e. the periodic response at matching fields, are independent of the precise physical origin of the periodic pinning potential as are magnetic dots and the roughness of the film, as described in reference 24 and or the magnetic state of the dots as discussed in reference 25.

We have carried out measurements using different experimental techniques in a broad temperature range. In the present study, several experimental techniques overlap in the same temperature interval. We found that vortex pinning is present in all the temperature range whereas Little-Parks oscillations are only operational in a narrow temperature range where the coherence length is longer than the inter dot distance. We have also established the threshold for Little-Parks oscillations and are able to separate in the periodic responses the contributions arising from pinning or from Little-Parks mechanisms. We conclude that in the present hybrid systems, i. e. array of nanomagnets in thin films, the periodic responses are mainly governed by pinning with a small contribution from the Little-Parks effect in a narrow temperature interval close to $T_c$. This very important conclusion directly impacts on theoretical models used to understand periodic response of hybrid superconducting-magnetic arrays.

The paper is organized as follows: in the next section sample fabrication and experimental methods are described; afterward we present the experimental data, beginning with magnetic measurements which include ac susceptibility and magnetization and following with the transport measurements, which include magnetoresistance and critical current; this section is closed with the analysis and the discussion of the experimental results. Finally, the paper ends with a summary section.

## 2. Experimental methods

The hybrid samples are fabricated of arrays of magnetic Ni nanodots embedded in superconducting Nb films. The arrays were grown on Si (100) substrates using electron beam lithography and lift-off in combination with dc magnetron sputtering. The Ni nanodots dimensions are 200 nm diameter and 40 nm thickness and are arranged on a square lattice of side a = 400 nm. The magnetic dots are in the magnetic vortex state, as observed in Ni nanotriangles of similar dimensions [26] and in direct electronic contact with the superconductor. A 100 nm thick Nb film was deposited by dc magnetron sputtering on top of the arrays, forming the magnet/superconductor hybrids, i. e. Ni magnetic dots covered by a Nb superconducting film. Two samples were measured. The first one was used for magnetic measurements. Total dimension of this array is 3 mm x 3.5 mm. The second one was patterned to a cross-shaped bridge with 40 μm x 40 μm bridge dimension for transport measurements.

We have applied several experimental techniques to probe superconducting behaviour in different temperature intervals. In this fashion, a much wider temperature range than usually reported in the literature, could be investigated (see for example reference 7 and references therein). These experimental techniques are: i) ac susceptibility as a function of magnetic field at constant temperature, $\chi_{ac}(H)$; ii) zero field cooled (ZFC) dc magnetization loops at constant temperature, M(H); iii) (I,V) curves at different temperatures and magnetic fields which provide the critical currents *vs.* applied magnetic fields, $I_c(H)$; and, iv) dc magnetoresistance which allows studying dissipation *vs.* applied magnetic fields, R(H). In figure 1 we have plotted the upper critical fields *vs.* temperature showing the temperature intervals where the different techniques span.

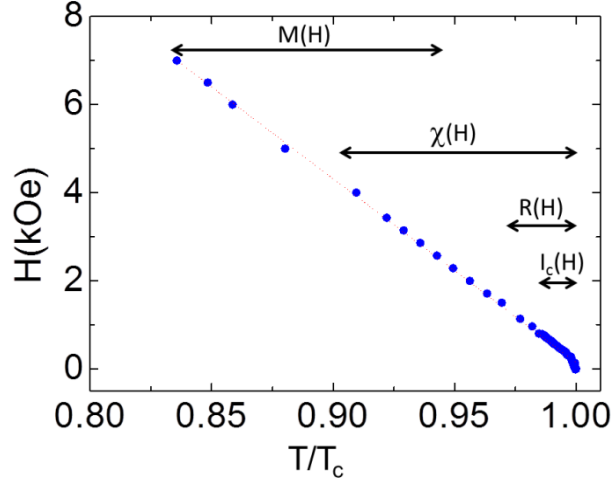

**Figure 1.** Upper critical field (H) vs. normalized temperature to the zero field critical value ($T_c$). Arrows indicate the temperature intervals where the different superconducting properties have been measured. (Colour online).

Magnetic measurements were performed in a Quantum Design MPMS superconducting quantum interference device (SQUID) magnetometer with dc and ac magnetic fields applied normal to the sample surface. Ac susceptibility as a function of temperature, $\chi_{ac}(T)$, was measured in field cool (FC) conditions for different applied dc fields. Ac susceptibility, $\chi_{ac} = \chi´ + j\chi´´$ is the first harmonic component of the Fourier transform of the time dependent magnetization. $\chi´$ provides the shielding capability and $\chi´´$ the ac losses. For perfect shielding, $\chi´ = -1$ and its absolute value decreases (tends to zero) as the ac field penetrates into the sample due to increased vortex lattice (VL) mobility. Constant temperature zero field cooled (ZFC) dc magnetization loops M(H), were measured to address issues related to relaxation phenomena. The characteristic time in M(H) measurements is approximately 60 sec, while $\chi_{ac}$(H) has a characteristic measurement time $\tau = 1/f$ [27]. Therefore, if relaxation dominates the behaviour, low frequency ac experiments are needed for a comparison to dc magnetization. However, for these low $T_c$ superconducting samples we found no significant difference for frequencies between f =1 Hz and 1 kHz. The data presented here is for an ac amplitude $h_{ac}$ = 1 Oe and f = 1 kHz.

A commercial Helium cryostat with variable temperature insert and a superconducting solenoid is used for the magnetotransport measurements. The magnetic field is applied normal to the sample plane. Vortices are driven by dc currents injected in the patterned cross-shaped bridge which allows standard four points resistance measurements.

## 3. Experimental results and discussion

The magnetic behaviour of superconductors is one of the richest and most interesting subjects in the field. Here we focus on the low field magnetic behavior. Constant temperature zero field-cooled (ZFC) dc magnetization curves, M(H), are plotted in figure 2(a). Low field magnetic instabilities arising from thermomagnetic effects [28] lead to flux avalanches which have also been reported in Pb [29] and in Nb films [30]. The matching features appear as "shoulders" in M(H) at regular magnetic field intervals which are enhanced as T is reduced.

A comparison of the ac susceptibility (figure 2(b)) and M(H) (figure 2(a)) shows clearly observable periodic features appearing at the same fields. At low temperature, intrinsic pinning competes with the artificial periodic pinning landscape, therefore matching periodic features become more evident with temperature (see for example reference 22). However, comparing both techniques, the highest temperature where modulations in M(H) measurements are observed is T ~ 0.95$T_c$, while in χ'(H) matching effects are observed up to 0.99$T_c$. On the other hand, the low temperature limit for which we observe matching is T = 0.78$T_c$ in M(H) and T = 0.90$T_c$ in χ' (H) due to the signal and noise levels.

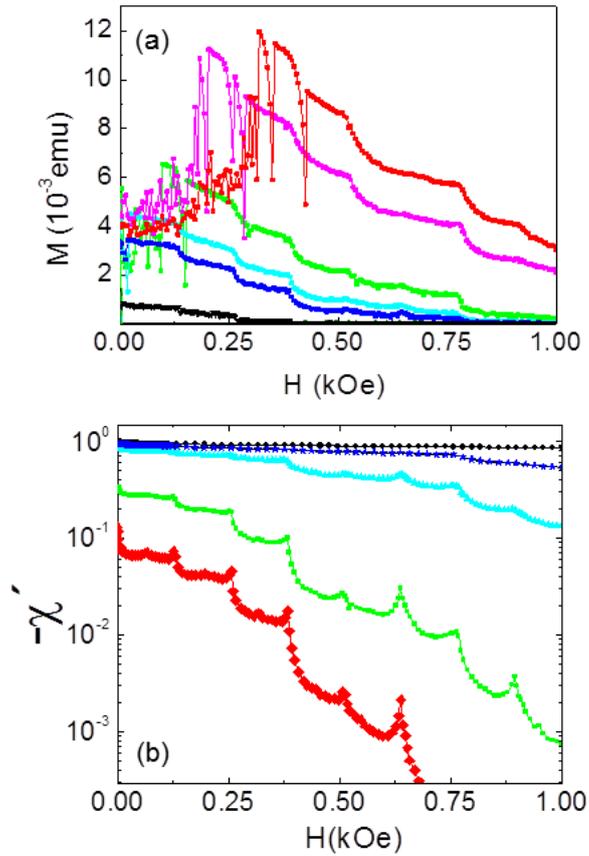

**Figure 2. (a)** Magnetization (M) as a function of the magnetic field (H) at different temperatures: 0.78$T_c$ (red), 0.81$T_c$ (pink), 0.86$T_c$ (green), 0.89$T_c$ (cyan), 0.90$T_c$ (blue) and 0.94Tc (black). Note that periodic "shoulders" observed in the magnetization indicate increase in critical current density matching effects. **(b)** Real part (shielding capability) of ac susceptibility χ' as a function of the magnetic field (H) at different temperatures: 0.99$T_c$ (red), 0.98$T_c$ (green), 0.94$T_c$ (cyan), 0.90$T_c$ (blue) and 0.79$T_c$ (black). The high temperature spikes at the matching fields become shoulders at lower temperature. (Colour online).

Figure 3(a) and 3(b) shows magnetoresistance and critical current data respectively. Sharp, well defined periodic minima or maxima are clearly observed in the magnetoresistance or critical current as a function of the applied magnetic fields.

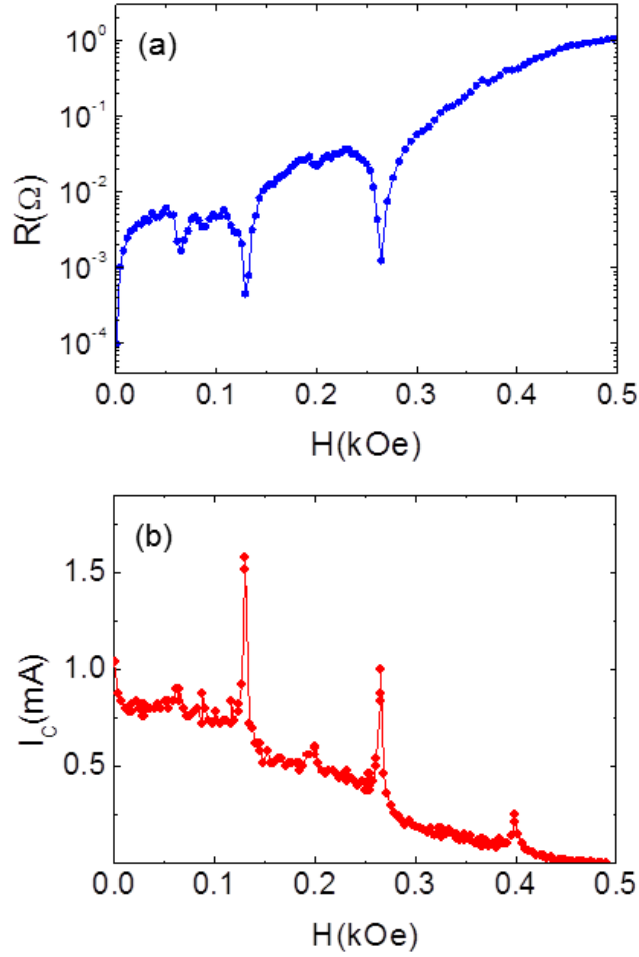

**Figure 3.** Magnetotransport measurements at $0.99T_c$. **(a)** Magnetic field dependence of the resistance (R) for an applied current I=1 mA. **(b)** Critical current ($I_c$) as a function of the magnetic field using a voltage criterion of 12.5 mV/mm.

The interpretation of these periodic features has been addressed by much work in the past [7]. In brief, the first or main minimum (maximum) appears at a magnetic field $B = (\Phi_0/S)$, where S is the unit cell area of the pinning array and $\Phi_0 = 20.7$ G μm$^2$ is the magnetic flux quantum. Other minima (maxima) occur at matching fields $B_n = n (\Phi_0/S)$, where n > 1 is an integer number.

These experimental results show that the commensurability effects appear at the same periodic magnetic fields independently of the experimental techniques employed. All experiments presented here show that commensurability effects have been detected in a wide temperature range, with different techniques which probe static as well as dynamic vortex behaviour. Matching effects between the periodic pinning sites and the vortex lattice is the only interpretation which provides a simultaneous explanation for the periodic field response in magnetic and transport measurements. Moreover, the periodic M(H) features appear as shoulders and not cusps as commonly expected for a wire network [15].

The definite argument which supports vortex pinning as the origin of the oscillatory behaviour is the comparison between the size and periodicity of the array with the coherence lengths (ξ). Typical wire network behaviour appears when ξ becomes of the same order or larger than the width (W) of the superconducting "stripes" between dots. This W is given by the difference between the dot periodicity (a), and the dot diameter (d), i.e. W = a − d = 200 nm ≤ $\xi(T)$. In our case, wire network regime appears when $\xi(T) \geq 200$ nm.

The coherence length $\xi(0)$ is determined from a fit of the measured linear critical field vs. temperature (figure 1) as given by the well-known expression for the upper critical field, $H_{c2}(T) = \Phi_0/2\pi\xi^2(T)$. This gives the dirty limit $\xi(0) = 9$ nm in good agreement with many other measurements of $\xi(0)$ in similar samples. In this work, the shortest attained coherence length is obtained at $0.78T_c$ (the lowest temperature studied here), which corresponds to a coherence length $\xi(0.78T_c) \sim 19$ nm. This is much smaller than the separation between the dots (W = 200 nm) which is clearly incompatible with the wire network description.

However, since the coherence length diverges at the superconducting transition temperature, it may become of the same order as the width of the superconducting "stripes" between the dots. This therefore can give rise to the Little-Parks (LP) effect for T(H) very close to the critical temperature $T_c(H=0)$. The crossover temperature, above which the superconducting wire network (SWN) behaviour appears, can be obtained from a simple consideration. The LP crossover should occur at a temperature that satisfies that $1.84\xi(T) = W$. From this, we obtained that a crossover should be obtained at $T = 0.993T_c$ as indicated by the dashed line in figure 4. This figure 4 shows that, at temperatures above the crossover, maxima in the critical temperature are obtained at the matching fields. They follow a parabolic background [17] typical of a superconducting wire network (SWN). So, at these high temperatures, the hybrid systems mimic a superconducting wire network. At temperatures below the SWN crossover, the upper critical field follows a linear T dependence, as expected for superconducting films in perpendicular magnetic fields (see figure 1).

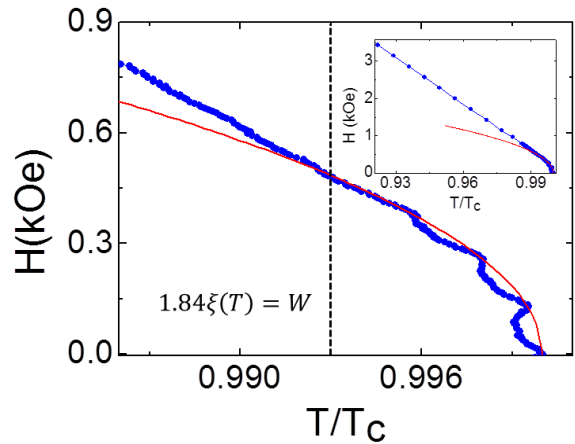

**Figure 4.** Upper critical field (H) vs normalized temperature to the zero field critical value ($T_c$). Red line is the parabolic fitting of the maxima obtained in $T_c(H)$ at the matching fields. Inset shows a higher temperature range. For temperatures lower than the crossover to the LP effect, a linear dependence characteristic of thin film behaviour is obtained. The critical temperatures are obtained following the same $0.5R_N$ criterion used in [18]. In our case $T_c(H=0) = 8.44$ K. (Colour online).

This crossover has been extensively studied and reported, however, the role of vortex pinning in this temperature range has not been taken into consideration. Following, we discuss the role of pinning at high temperatures where SWN is unambiguously observed. Giroud et al.[31] found in weakly coupled $AlO_x$ wire networks, periodic oscillations together with a substantial broadening of the superconducting transitions in magnetic fields. The oscillation period is in very good agreement with the expected values calculated from the array period in the Al network. Interestingly Giroud et al. [31] observe a periodic broadening of the resistive transition in a magnetic field, with a larger shift in the low resistance tail. This is the same behaviour reported by Patel et al. [17] and found by us. The lowest amplitude is observed at $R/R_N = 0.9$ and the largest at $R/R_N = 0.1$. This is a clear indication that pinning and vortices play a role in the oscillations amplitude. To clarify this behaviour we measured the broadening parameter $b = T(0.9\ R_N) - T(0.1\ R_N)$ as a function of the applied magnetic field (figure 5).

Figure 5 shows the following remarkable results: i) the smallest broadening appears at the matching conditions, either above or below the crossover. This is a clear indication that pinning is operational above and below the crossover, because at matching pinning is stronger than out of matching. ii) When the sample is in the thin film regime (high magnetic fields in figure 5) the average value of b is higher than in the region of lower magnetic fields when the hybrid mimics a SWN sample.

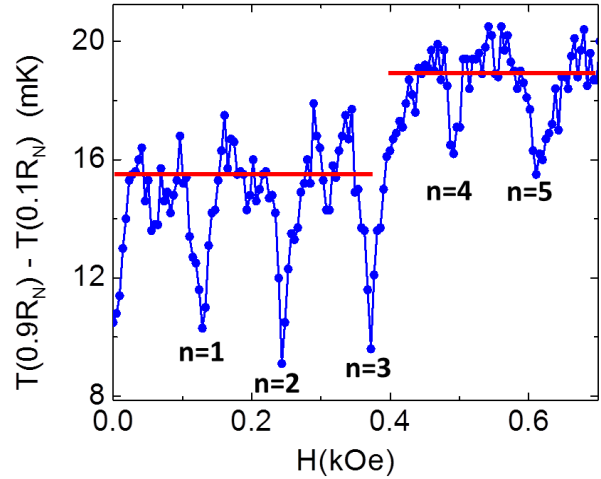

**Figure 5.** Parameter b (width of the resistive superconducting transition) as a function of the perpendicular applied magnetic field. Minima in the width are observed at the matching fields (n indicates the index of the matching field $H = n\ H_{match}$). An increase in the transition width is observed in the crossover from the superconducting wire network to the thin film regime (n>3) (red lines are guides to the eye). (Colour online)

These experimental data allow investigating the influence of pinning on the Little–Parks (LP) effect, when both mechanisms coexist. The dirty limit theoretical expression [32] for the maximum LP oscillations is given by

$$\Delta T_C(H) = 0.73\ T_C \frac{\xi_0 \ell}{R^2} \left(n - \frac{\phi}{\phi_0}\right)^2 \qquad (1)$$

where $\ell = 2.84$ nm, $\xi_0 = 39$ nm, R = 100 nm is the radius of the dots. These oscillations, represented with a black line in figure 6, imply a maximum theoretical oscillation of

$\Delta T_{c,th}(H)=17$ mK. The red dots in figure 6 represent the critical temperature oscillations ($\Delta T_c$(H)) extracted from the experimental data (blue dots in figure 4) by subtracting the parabolic background (red line in figure 4). The experimental maximum amplitude is $\Delta T_{c,exp}(H) = 7$ mK. These results show an unambiguous decrease in the LP oscillations in this type of samples where vortex pinning is also present.

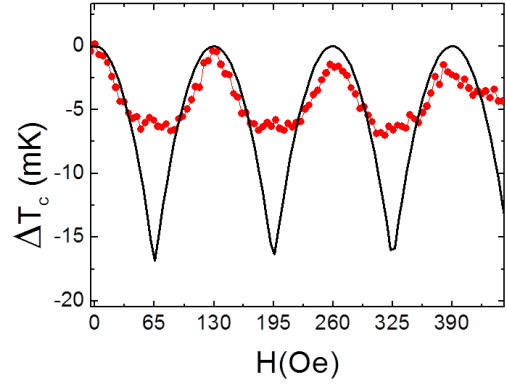

**Figure 6.** Red dots show $\Delta T_c(H)$ obtained by subtracting the parabolic background (red line in figure 4) from the experimental measurements (blue dots in figure 4). Black line shows the theoretical Little-Parks oscillations in the critical temperature (see text). The maximum amplitude obtained experimentally is smaller than the estimated theoretical value. (Colour online).

Thus figure 5 and figure 6, show two effects produced by the pinning. First, minima appear in the superconducting transition width at matching conditions. Second, the LP oscillations are diminished, increasing the minimum critical temperature in out of matching conditions as the vortex pinning slows down the vortex motion. It is important to note that the effects showed in figures 5 and 6 arise naturally from models which have as an essential ingredient vortex pinning and to the best of our knowledge cannot be explained using the Little-Parks effect.

**4. Conclusions**

Oscillations in the magnetic and transport properties in a broad temperature interval are observed in superconducting thin films grown on top of arrays of magnetic dots. These periodic oscillations appear at matching magnetic fields which are set by the array geometry. This behaviour is due to commensurability of the vortex lattice with the array of pinning centers. At temperatures close to the critical temperature the magnitude of the coherence length can become of the same order as the size (dimension) of the channels in between the dots. Because of this, the sample can mimic a weakly coupled superconducting wire network. This behaviour produces experimentally observable Little-Parks oscillations in the critical temperature. Remarkably, the pinning mechanism coexists with the wire network regime as shown by periodic broadening of the resistive transition.

In summary, in superconducting films with array of magnetic dots the following experimental facts are observed: i) the magnetic dot arrays decrease the resistive transition broadening (T(0.9 $R_N$) - T(0.1$R_N$)) at commensurability between the vortex lattice and the magnetic array; ii) In the

SWN regime the resistive transition broadening (T(0.9$R_N$)-T(0.1$R_N$)) decreases; iii) The experimental amplitude of the critical temperature oscillations $\Delta T_c(H)$ is smaller than the estimated theoretical value obtained from the LP model. From these experimental results, we conclude that both LP and pinning mechanisms coexist in this type of samples in a small temperature range close to $T_c$. However, in contrast to superconducting films with arrays of nanoholes, the main contribution is vortex pinning that is maintained in a broad temperature range.

**Acknowledgements**

We thank the support from Spanish MINECO grants FIS2008-06249 (Grupo Consolidado), Consolider CSD2007-00010 and CAM grant S2009/MAT-1726. The magnetism aspects of this work were supported by the Office of Basic Energy Science, U.S. Department of Energy, under Grant No. DE FG03-87ER-45332 and Argentina UBACyT 661 and PICT 2008 No 753.